\documentclass[twocolumn,aps,prl]{revtex4}
\usepackage{epsfig,graphicx}
\begin{document}

\title{\bf  Atomic-scale perspective on the origin of attractive step interactions on Si(113) }
\author{Cristian V. Ciobanu, Dhananjay T. Tambe and Vivek B. Shenoy }
\affiliation{Division of Engineering, Brown University,
Providence, RI 02912}
\author{Cai-Zhuang Wang and Kai-Ming Ho}
\affiliation{ Ames Laboratory - USDOE and Department of Physics,
Iowa State University, Ames IA 50011}

\date{\small \today}
\begin{abstract}
Recent experiments have shown that steps on Si(113) surfaces
self-organize into bunches due to a competition between long-range
repulsive and short-range attractive interactions. Using empirical
and tight-binding interatomic potentials, we investigate the
physical origin of the short-range attraction, and report the
formation and interaction energies of steps. We find that the
short-range attraction between steps is due to the annihilation of
force monopoles at their edges as they combine to form bunches.
Our results for the strengths of the attractive interactions are
consistent with the values determined from experimental studies on
kinetics of faceting.
\end{abstract}
\maketitle


Self-assembly of steps is an important way to achieve surface
patterning at length scales where the usual lithographic
techniques are not applicable. On certain vicinal semiconductor
surfaces, the equilibrium bunching of steps leads to the formation
of periodic hill-and-valley patterns (or grooves), that can serve
as templates for growing regular arrays of one-dimensional (1D)
nanoscale devices like quantum wires. Experimental work in recent
years
\cite{mochriePRL,mochrieGrooveSS,mochrieSS,sudohPRL,zippingLetter}
has revealed that such groove structures are formed on vicinal
Si(113), where the flat (113) surface coexists with step-bunched
regions over a wide range of temperatures and misorientation
angles.

In a series of experiments, Mochrie and coworkers
\cite{mochriePRL,mochrieGrooveSS,mochrieSS} have studied the
orientational phase diagrams for stepped Si(113) surfaces miscut
towards several crystallographic orientations. In all the cases,
their studies strongly suggest that, at low temperatures, the
steps gather in bunches as a result of attractive step-step
interactions. Independent experimental work on the fluctuations of
the single and multiple-height steps and the kinetics of faceting
transitions by Sudoh and coworkers \cite{sudohPRL,zippingLetter}
have confirmed the presence of short-range attractive interactions
between steps. Furthermore, their analysis of the step-zipping
process has provided quantitative estimates for the magnitude of
the attractive interactions \cite{zippingLetter}. The presence of
such interactions has been assumed in theoretical models that show
the orientational phase diagram \cite{mochriePRL} can be explained
in terms of competing short-range attractive and long-range
repulsive step interactions \cite{lassig,vbsPRL}. While much
progress has been made in understanding the self assembly of steps
on Si(113) surfaces, a microscopic analysis of the step
interactions has not been attempted and the physical origin of the
short-range attraction has not been convincingly explained.

%

In this letter, we investigate the interactions between steps on
Si(113), with the goal of understanding the origin of the
effective short-range attraction between steps. Describing each
step as a pair of equal and opposite force monopoles located at
the step edges, we find that the short-range attraction between
steps is due to the cancellation of force monopoles that belong to
adjacent steps in a step bunch. This observation is confirmed by
investigating the scaling of the step formation energies and
interaction strengths with the step height. Noting that the
structure of the steps is determined by the direction of the
miscut, we consider here Si(113) surfaces miscut towards
$[\overline{3}\overline{3}2]$, so that our estimates for the
magnitudes of step interactions can be directly compared with the
recent experiments of Sudoh {\em et al.} \cite{zippingLetter}.

\begin{figure}
  \begin{center}
   \includegraphics[width=3.0in]{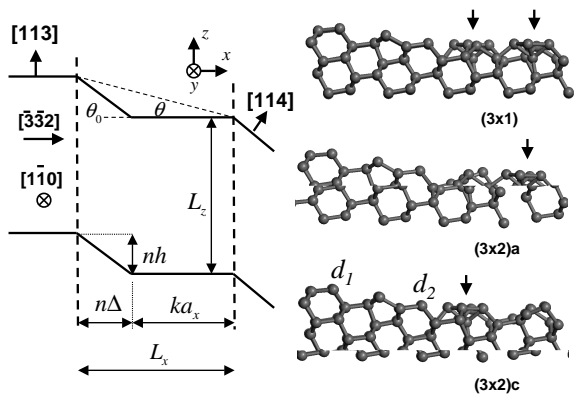}
  \end{center}
\caption{(a) Geometry of a typical stepped Si(113) surface with
the simulation cell enclosed between thick dashed lines.
Different step structures with $n=k=1$
(miscut $\theta=3.2^o$) are shown on the right. Arrows mark the
positions of the interstitial atoms that stabilize the terraces
\cite{dabr}. For the (3$\times$2) surfaces, there are two possible
ways of creating a step: the interstitial rows on the {\em lower}
terrace can be located either away from the step, (3$\times$2)a,
or {\em close} to the step, (3$\times$2)c. The symbols $d_1$ and
$d_2$ label the dimers on the (114) step. } \label{supercell}
\end{figure}

The slab geometry that shows the Si(113) terraces and types of
steps studied in the present work is given in
Fig.~\ref{supercell}. Steps run along the $[1\overline{1}0]$
direction and consist of (114) nanofacets \cite{zippingLetter},
with a ($2\times 1$) reconstruction \cite{erwin114}. For the (113)
surface we consider both the (3$\times$1) and the (3$\times$2)
structures \cite{dabr} that are stabilized by 
interstitial atoms.
We distinguish two possible
structures for the (3$\times$2) case, depending on the position of
the interstitial rows on the terraces with respect to the step.
When the interstitials on the {\em lower} terrace are {\em away}
from the step or {\em close} to the step, we call them the
(3$\times$2)a and the (3$\times$2)c structures, respectively
(refer to Fig.~\ref{supercell}).

The dimension of  the simulation cell in the $x$-direction, $L_x$,
is determined by the number $k$ of Si(113) unit cells and the
number $n$ of Si(114) unit cells that constitute the step: $L_x=
n\Delta + ka_x $, where $a_x=12.73$\AA\  and $\Delta=16.21$\AA\
are, respectively, the periodic lengths of the (113) and the (114)
surfaces along $[\overline 3 \overline 3 2]$
(Fig.~\ref{supercell}). The periodic length in the $[1 \overline 1
0]$ direction $L_y$ is the smallest length compatible with the
periodicity of both the Si(113) and the Si(114) structures,
$L_y=23.04$\AA. The thickness of the slab $L_z$ was chosen to be
$\sim$150 \AA. 
To simulate the vicinal geometry, we use shifted periodic boundary
conditions \cite{farid} in which 
the shift along [113] is determined by the step height $nh$, where
$h=1.64$\AA \  is the height of a single step.

The energies of the slabs in  Fig.~\ref{supercell} are computed
using empirical and tight-binding (TB) models for the atomic
interactions \cite{abinitio}. Empirical calculations are carried
out with the Tersoff potential T2 \cite{tersoff2} as it reproduces
the structure, surface stress and relative energy of the Si(113)
and Si(114) reconstructions reasonably well. While the
tight-binding methods are more accurate, they are usually
impractical for ledge energy calculations since they require slabs
with several thousands of atoms to capture the long-range
interactions between steps. To circumvent this problem, we have
developed a multi-scale approach where the tight-binding model is
coupled to an  empirical potential as described in the following
paragraph.

\begin{figure}
  \begin{center}
   \includegraphics[width=3.2in]{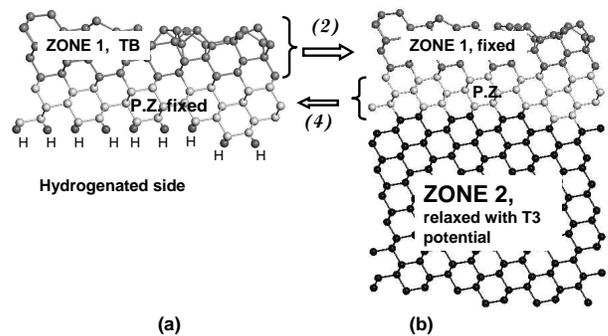}
  \end{center}
\caption{Relaxation scheme employed to compute ledge energies
using a coupling of the tight-binding (TB) method \cite{czwJPCM}
with the T3 potential \cite{tersoff3}. Zone 1 (thin) consists of a
few surface layers that are relaxed using TB (a), but are kept
fixed during the relaxation of the thick slab (b). Zone 2 (thick)
represents the deeper layers, which interact with the surface
layers through a 
padding zone P.Z. The arrows show how atomic coordinates are
updated during stages {\em (2)} and {\em (4)} of the procedure
described in text.} \label{matchingslabs}
\end{figure}

The key idea of the coupling scheme is to handle the long-range
elastic fields produced by steps using an empirical potential
whose elastic constants are very close to their TB counterparts;
this would ensure that the elastic fields in the hybrid scheme
would not be very different from the fields obtained in the case
where all the atoms were treated using TB. We use a charge
self-consistent TB method, whose superior transferability has been
studied in detail in \cite{czwJPCM}. This method is coupled with
the Tersoff (T3) potential \cite{tersoff3} using the scheme shown
in Fig.~\ref{matchingslabs}. We divide the simulation cell into
three regions: (1) a thin surface region where the complex bonding
topologies in the vicinity of the step-edge are treated with the
TB method, (2) a thick "bulk-like" region where all the atoms with
tetragonal bonding environments are treated with the T3 method and
(3) a padding zone which is used to communicate atomic
displacements between the first two regions. The atoms in the
simulation cell are relaxed using the following procedure (refer
to Fig.~\ref{matchingslabs}): {\em (1)} relax the thin slab (Zone
1, Fig.~\ref{matchingslabs}(a)) using TB, while keeping the
padding zone fixed; 
{\em (2)} update the atomic coordinates of Zone 1 in
Fig.~\ref{matchingslabs}(b) with the values obtained after the TB
relaxation of Zone 1 in Fig.~\ref{matchingslabs}(a);
{\em (3)} relax the thick slab and the padding zone (Zone 2 and
P.Z. in  Fig.~\ref{matchingslabs}(b)) using
the T3 potential while keeping the atoms in Zone 1 fixed; 
{\em (4)} update the coordinates of the P.Z. atoms in
Fig.~\ref{matchingslabs}(a) with the values obtained
after the T3 relaxation  of P.Z. in Fig.~\ref{matchingslabs}(b); 
{\em (5)} repeat steps {\em (1)} through {\em (4)} until energy
convergence is achieved \cite{elsewhere}.

%

Once the total energy $E$ of the slab is calculated, step
interactions can be obtained from the ledge energy \cite{farid}
\begin{equation}
 \lambda (L_x) \equiv  (E-Ne_b-\gamma_{113}L_xL_y)/L_y, \label{lambda}
\end{equation}
where $e_b$ is the bulk cohesion energy per atom \cite{ebulk}, $N$
is the number of atoms in the cell in Fig.~\ref{supercell}, and
$\gamma_{113}$ is the surface energy of Si(113).
\begin{figure}
  \begin{center}
   \includegraphics[width=3.0in]{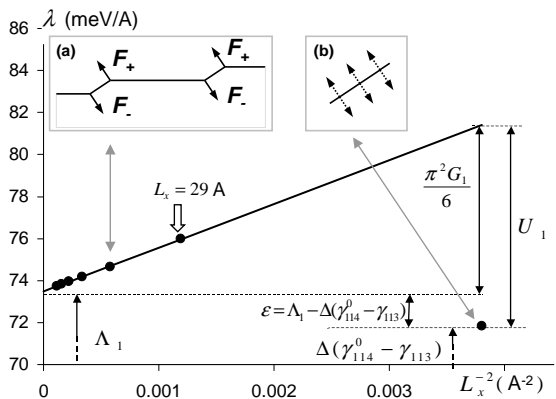}
  \end{center}
\caption{Ledge energy $\lambda$ of the single-height (114) steps
separated by (3$\times$2)c terraces, plotted as a function of
$1/L_x^2$. The points corresponding to separations $L_x>
\Delta=$16.21\AA \ are fit to a straight line, $\lambda =
\Lambda_1 + \pi^2 G_1 \Delta^2/6L_x^2$. At the smallest separation
$\Delta$, the ledge energy falls {\em below} the straight line,
indicating a short-range attraction
$U_1=\pi^2G_1/6+\Lambda_1-\Delta(\gamma_{114}^0-\gamma_{113})$
\cite{114}. The inset (a) shows an array of steps (simulated via
periodic boundary conditions) where each step is represented
schematically as a pair of force monopoles $F_+$ and $F_-$. When
$L_x=\Delta$, there are no terraces between steps and the
monopoles corresponding to adjacent steps cancel each other, as
shown in inset (b). }\label{fitlambda}
\end{figure}
Using the fact that the steps interact through long-range dipolar
fields \cite{Mdipoles}, we have analyzed the short-range
interactions by fitting the ledge energies in Eq.~(\ref{lambda})
to the form \cite{vbsPRL}
\begin{equation}
\lambda (L_x=n\Delta+ka_x) = \Lambda_n +G_n\frac{\pi^2
\Delta^2}{6L_x^2} - U_n \delta_{k0}, \label{eqfitlambda}
\end{equation}
where $\Lambda_n$ is the formation energy per unit length of a
step of height $nh$, $G_n$ is the strength of the long-range
dipolar interactions, and $\delta_{k0}$ denotes the Kronecker
delta symbol, where $k$ is the number of (113) unit cells on the
terrace (Fig.~\ref{supercell}). The energy offset $U_n$ that
appears in Eq.~(\ref{eqfitlambda}) describes the effective
interactions at the smallest possible spacing between the steps,
$n\Delta$ (note that the (113) terraces are absent in this case).
A positive (negative) value of $U_n$ indicates the presence of a
short-range attraction (repulsion) between steps. Our results for
different terrace  and step structures are discussed below.

%

We first focus attention on the vicinal surfaces made of
single-height steps and illustrate the general features of the
step interactions by considering the energetics of the
(3$\times$2)c steps obtained from the T2 potential. In
Fig.~\ref{fitlambda}, we plot the ledge energy as function
$1/L_x^2$ ($L_x$ is the step separation) and provide the fitting
parameters $\Lambda_1$, $G_1$ and $U_1$ defined in
Eq.~(\ref{eqfitlambda}). If the dipolar description of step
interactions is assumed to hold over the entire range of step
separations (as implied by Eq.~(\ref{eqfitlambda})),
Fig.~\ref{fitlambda} shows the presence of a short-range
attraction ($U_1$), since the point corresponding to $L_x =
\Delta$ lies {\em below} the linear fit for the ledge energy. The
physical reason for the origin of this attraction will be
discussed next.

We start by noting that the Si(113) and Si(114) facets are
atomically compatible, in the sense that they can meet to form
edges parallel to [1$\overline{1}$0] without any changes in their
individual bonding topologies. Subsequently, the elastic fields of
a (114) step (or nanofacet) can be expressed in terms of a dipole
which consists of a pair of equal and opposite force monopoles
($F_{+},\ F_{-}$)\cite{Mgrooves} at the edges of the step, as
shown in inset (a) of Fig.~\ref{fitlambda}. While such steps are
expected to show inverse-square repulsive interactions at large
separations, we find that this behavior holds for separations  as
small as 29\AA \ (this point is marked in Fig.~\ref{fitlambda}).
When the steps are closer than 29\AA, they lose their
individuality forming a smooth Si(114) surface: at a separation of
$L_x=\Delta$ there is a cancellation of the monopoles at the
step-edges as illustrated in inset (b) of Fig.~\ref{fitlambda}.
The absence of dipoles with mutual repulsive interactions is the
reason for the presence of an effective short-range attraction
between steps. Further evidence for this argument is obtained by
computing the interactions of multiple-height steps. We will focus
on such calculations after we discuss the interaction parameters
for other types of steps given in Fig.~\ref{supercell}.
\begin{table}
\begin{tabular}{ l l c c c c c}
\hline \hline Method & Structure \  & \ $U_1$\ & \ $G_1$\     & \ $U_1/G_1$  & $ \Lambda_1$ & $\varepsilon$ \\ 
\hline
\           &(3$\times$1)in      & 61.68  &  18.84 &   3.27  &   108.94  & 26.05 \\ 
{\bf \em TB-T3} & (3$\times$1)out     & 60.12  &  13.65 &   4.40  &    110.28 & 33.08 \\ 
\           &(3$\times$2)c, in    & 62.6   &  8.69  &   7.2  &    66.48  & 45.20 \\ 
\hline
\            &(3$\times$1)       & 8.74 &  10.45   & 0.84 &   173.59  & -8.44 \\ 
{\bf \em T2} & (3$\times$2)a      & 0.22 &  3.70    & 0.06 &   65.98   & -5.86 \\ 
\            &(3$\times$2)c      & 9.30 &  4.81    & 1.93 &   73.48   & 1.64  \\ 
\hline
{\bf \em Exp.} \cite{zippingLetter}&(3$\times$2)  &   &   &   &  57 &  13--15 \\
\hline
 \hline
\end{tabular}
\caption{Interaction strengths and step formation energies for
single-height steps computed using the multiscale
tight-binding/Tersoff scheme (TB-T3) and the  T2 potential
\cite{tersoff2}.
The labels "in" and "out" denote the relative tilting (in- and
out-of-phase) of the dimer rows $d_1$ and $d_2$ shown in
Fig.~\ref{supercell}. The last column shows the quantity
$\varepsilon \equiv
\Lambda_1-\Delta(\gamma_{114}^0-\gamma_{113})$, which allows for a
direct comparison with the experimental results (Exp.
\cite{zippingLetter}). All quantities (except the ratio $U_1/G_1$)
are given in meV/\AA. } \label{LUG}
\end{table}

Using both the T2 potential and the TB-T3 method, we find the
existence of short-range attractive interactions between steps for
all the reconstructed surfaces considered here. However, the
magnitudes of these interactions depend strongly on the structures
of the terraces and steps (refer to Table~\ref{LUG}). Since the TB
method allows for the tilting of the dimers $d_1$ and $d_2$ on the
step (marked in Fig.~\ref{supercell}), we have reported the
results for the cases where the tilting patterns are in-phase and
out-of-phase \cite{restrict}. The magnitude of the attractive
interaction $U_1$ is $\approx60$meV/\AA \ and $\approx10$meV/\AA \
for the TB-T3 method and T2 potential, respectively. This
discrepancy between the values of $U_1$ computed with TB-T3 and
with T2 is due to the differences in step formation energies
obtained with the two methods, as well as to the sensitivity of
$U_1$ with respect to changes in $(\gamma_{114}^0-\gamma_{113})$
(refer to Fig.~\ref{fitlambda}). For both T2 and TB-T3, the
repulsion strength $G_1$ is of the order of 10meV/\AA, while the
formation energy $\Lambda_1$ is of the order of 100meV/\AA \ (see
Table~\ref{LUG}). In order to compare our results with the
experimental ones \cite{zippingLetter}, we note that Sudoh {\em et
al.} \cite{zippingLetter} report the quantity $\varepsilon \equiv
\Lambda_1-\Delta(\gamma_{114}^0-\gamma_{113})$, i.e. the
difference between the step formation energy $\Lambda_1$ and the
excess energy of the Si(114) facet per length of the step.
Table~\ref{LUG} shows that the values for $\varepsilon$  obtained
with the TB-T3 method are of the same order of magnitude as the
estimates of Sudoh {\em et al.} \cite{zippingLetter}. Closer
agreement could be achieved if the long-range interactions are
included in the analysis of kinetics of step-zipping reported in
\cite{zippingLetter}. Also, future experiments aimed at
determining the structure of the step ($a$ or $c$) could verify
the trends given in Table~\ref{LUG}.

We will now focus on the formation energy and interactions of
steps with height $nh$ ($n$-bunches) for $n=2,\ 3$. Because of the
larger size of the simulation cells, we have used only the T2
potential \cite{tersoff2} for this analysis. In the zero
temperature limit, the formation energy of an $n$-bunch can be
expressed as the sum of the formation and interaction energies of
the steps within the bunch \cite{vbsPRL}:
\begin{equation}
 \Lambda_n  =n \Lambda_1-U_1(n-1)+G_1\sum_{i=1}^{n-1}
\frac{n-i}{i^2}. \label{EDL-lambda}
\end{equation}
Due to the presence of attractive interactions between
single-height steps, the formation energy of an $n$-height step is
expected to show sub-linear dependence of $n$, $\Lambda_n <
n\Lambda_1$ \cite{vbsPRL}. This prediction is indeed borne out by
the atomistic results given in Table~\ref{LUG-h}. Using the
parameters $\Lambda_1,\ G_1, \ U_1$ from Table~\ref{LUG} for the
($3\times$2)c structure, we find that the bunch energies predicted
by Eq.~(\ref{EDL-lambda}) are within 1--2 meV/\AA \ of the values
for $\Lambda_n$ obtained in the present simulations.

Turning our attention to the step-step interactions, we find that
the repulsive interactions $G_n$ show a nearly quadratic
dependence of the step height $nh$, while the strength of the
attractive interaction $U_n$ shows a weak dependence \cite{weakUn}
on $n$ (refer to Table~\ref{LUG-h}). The quadratic behavior of
$G_n$ provides strong evidence to the idea regarding the
cancellation of the force monopoles associated with the adjacent
steps in an $n$-bunch. When such cancellation takes place, the
only remaining monopoles are associated with the {\em ends} of the
bunch. Consequently, each $n$-bunch behaves as a force-dipole with
a moment that is proportional to $nh$. Identical $n$-bunches
interact as dipoles \cite{Mdipoles}, with a repulsion strength
determined by the square of the dipolar moment of one $n$-bunch,
which results in the $n^2$ dependence of $G_n$.

\begin{table}
\begin{tabular}{ c c c c c c c}
\hline \hline  $nh$ \  & \ $ \Lambda_n $ \ &\ $n\Lambda_1$ \ & \
$\Lambda_n$,\ Eq.~(\ref{EDL-lambda})  & \ $G_n$ \ & \ $n^2G_1$\  &\  $ U_n $ \ \\
\hline $h$    &  73.48 &  73.48 & \       & 4.81   &  4.81  & 9.3   \\
$2h$          & 143.48 & 146.96 & 142.48  & 19.36  &  19.24 & 7.7   \\
$3h$          & 213.62 & 220.44 & 212.67  & 47.88  &  43.29 & 7.0  \\

\hline \hline
\end{tabular}
\caption{ Step-height dependence of  the formation energy
$\Lambda_n$ and interaction parameters $G_n,\ U_n$ for the
(3$\times$2)c structure, determined with the Tersoff potential T2
\cite{tersoff2}.
All quantities are given in meV/\AA .} \label{LUG-h}
\end{table}
%

%
In conclusion, we have shown that the effective short-range
attraction experienced by steps on Si(113) can be explained by the
cancellation of force monopoles associated with adjacent steps in
a step bunch. While numerical estimates for the interaction
parameters depend on the interatomic potential used, our physical
reasoning for the origin of attractive interaction is robust and
is consistent with experiments. A key point in understanding how
the atomic structure can lead to step-step attractive interactions
is the atomic compatibility of the two stable surface
orientations, Si(113) and Si(114): when nanofacets with these
orientations intersect forming a step edge, no bond breaking or
rebonding occurs. Recent experimental work \cite{takeguchi}
indicates that no rebonding occurs at the step-edges formed
between the Si(113) and Si(337) facets; our calculations show the
presence of attractive interactions between steps in this case,
further lending support to our physical picture. The difference
between the type of steps considered here, and the most common
steps on Si(001) surfaces, such as the SB and the DB steps
\cite{chadi} is that when steps on Si(113) merge they form a new
facet (e.g. Si(114)), while such atomically smooth facets are not
observed in the latter case due to significant rebonding at the
step-edges \cite{chadi}. Therefore, a simple cancellation of force
monopoles is not expected for the SB and the DB steps on Si(001),
and hence they do not show attractive interactions \cite{farid}.
Further theoretical and experimental studies aimed at elucidating
the dependence of the step-step interactions of the azimuthal
orientation will shed more light on the self-assembly of steps on
Si(113) miscut in an arbitrary direction.

%

Support from the MRSEC at Brown University (DMR-0079964), NSF
(CMS-0093714, CMS-0210095), and the Salomon Research Award from
the Graduate School at Brown University is gratefully
acknowledged. Ames Laboratory is operated by the U.S. Dept. of
Energy and by the Iowa State University (contract W-7405-Eng-82).


\begin{thebibliography}{30}
\bibitem{abinitio} Ab-initio studies to capture the elastic interactions
between steps on the (113) surfaces would require prohibitive
amounts of computational resources.
\bibitem{ebulk}In the case of the TB-T3 method, $e_b$ is the tight-binding
cohesive energy, which was chosen as the common scale for
expressing the total energy.
\bibitem{114} $\gamma^0_{114} = \gamma_{114}/\cos \theta_0$ (see Fig.~\ref{supercell}),
is the projected surface energy of Si(114).
\bibitem{restrict} In Table ~\ref{LUG}, we have included only the results for
cases that satisfy  $U_1/G_1 < 10$. Beyond this range, the theory
developed in \cite{vbsPRL} shows that the periodicity of the bunch
structures exceeds the experimental observations (100--1000 \AA)
\cite{mochriePRL,mochrieGrooveSS} by nearly 2 orders of magnitude.
\bibitem{weakUn} This result is consistent with the fact that $U_n$ is closely related to
the edge energy 
\cite{vbsPRL}, and should not depend on the number of steps within
the bunch.
\end{thebibliography}
\end{document}